\documentstyle[aps,twocolumn,epsfig]{revtex}

\begin{document}
\author{Zaher~Salman$^{1}$, Amit~Keren$^{1}$, Philippe~Mendels$^{2}$,
Valerie~Marvaud $^{3}$, Ariane~Scuiller $^{3}$,
Michel~Verdaguer$^{3}$, James~S.~Lord$^{4}$, and Chris~Baines$^{5}$}
\address{$^{1}$Technion-Israel Institute of Technology, Physics
Department, Haifa 32000, Israel \\ $^{2}$Laboratoire de Physique des
Solides, B\^{a}timent 510, UMR 8502 CNRS Universit\'{e} Paris Sud,
91405 Orsay, France \\ $^{3}$ Laboratoire de Chimie des M\'{e}taux de
Transition, URA CNRS 419 Universit\'{e} Pierre et Marie Curie, 75252
Paris, France\\ $^{4}$ISIS Facility, Rutherford Appleton Laboratory,
Chilton Didcot, Oxfordshire OX11 0QX, U.K.\\ $^{5}$ Paul Scherrer
Institute, CH 5232 Villegen PSI, Switzerland} 
\address{\vspace{2mm}\parbox{14cm}{\rm We
investigate the dynamical spin-spin auto-correlation function of the
isotropic high spin molecules CrCu$_{6}$ ($S=9/2$), CrNi$_{6}$
($S=15/2$) and CrMn$_{6}$ ($S=27/2$ ), using magnetization, $\mu $SR
and NMR measurements. We find that the field autocorrelation time
$\tau $ of the molecule's spin at zero and low fields is nearly
temperature independent as $T\rightarrow 50$ mK. The high temperatures
$\tau $ is very different between the molecules. Surprisingly, it is
identical ($\sim $ 10~nsec) at low temperature. This suggests that
$\tau $ is governed by hyperfine interactions.}}
\title{Dynamics at $T\rightarrow 0$ in the half-integer isotropic high spin
molecules}
\maketitle


High spin molecules (HSM) consist of clusters of metal ions, they are
ordered in a crystal lattice, and coupled by Heisenberg ferromagnetic
or antiferromagnetic interactions with coupling constant $J$, only
between spins $\vec{S}_{i}$ in the molecule. At low temperatures
$k_{B}T<J$ these spins lie parallel or anti-parallel to each other,
and the molecule is in its ground spin state, where
$\vec{S}=\sum_{i}\vec{S}_{i}$ is high, with quantum number $S$, and
$2S+1$ degeneracy. At even lower temperatures, $k_{B}T\ll J$ the
degeneracy can be removed by additional anisotropic interactions such
as the uniaxial term $DS_{z}^{2}$, or rhombic term
$E[S_{x}^{2}-S_{y}^{2}]$ etc. Experiments on the two most famous high
spin molecules Mn$_{12}$ \cite{ThomasNature96FriedmanPRL96} and
Fe$_{8}$ \cite {SangregorioPRL97,AubinJACS98} show that at low $T$ the
main interaction is, indeed, the uniaxial anisotropy, where up and
down spin states $S_{z}=\pm S=\pm 10$ are degenerate, and tunneling is
induced between these states by an additional term in the Hamiltonian
that does not commute with $S_z$. This quantum tunneling of the
magnetization (QTM) phenomenon has received considerable attention in
recent years, focusing mainly on the additional terms in the
Hamiltonian which are responsible for the tunneling \cite
{AubinJACS98,GaraninPRB97PolitiPRL95,ProkofevPRL98}. Nevertheless, the
theoretical picture is far from being clear and existing models are
controversial, and often contradict each other \cite{GargPRL98}. In
some cases, even qualitative understanding of the observed
experimental data is absent \cite{FriedmanJAP97}.

In this paper we investigate three simple HSM systems, which are
isotropic $(D,E=0)$. In these systems no tunneling is observed due to
the absence of the uniaxial term $DS_{z}^{2}$. However, spin dynamics
is observed even at very low temperatures ($T=50$ mK). Therefore, the
additional terms in the spin Hamiltonian could be probed directly in
these systems. Such a study can highlight the role of phonons
\cite{GargPRL98,VillainEL94}, dipolar interactions, and nuclear
fluctuations \cite{ProkofevPRL98} in a simple setup. In addition, it
could serve as a large $S$ limit for isotropic models usually applied
only for the $S=1/2$ case \cite{GargPRL98}, or it could serve as a
$D\rightarrow 0$ test case for anisotropic HSM models \cite
{PablosPRB97}.

We present an experimental investigation using three types of
measurements: magnetization, spin lattice relaxation ($T_{1}^{-1}$) of
muon spin ($\mu $SR), and of proton spin (NMR). Our molecules are
[Cr\{(CN)Cu(tren)\}$_{6}$ ](ClO$_{4}$)$_{21}$ \cite{Valerie01},
[Cr\{(CN)Ni(tetren)\}$_{6}$ ](ClO$_{4}$ )$_{9}$
\cite{Mallah95,Mallah96} and [Cr\{(CN)Mn(tetren)\}$_{6}$](ClO$_{4}$)$
_{9}$ \cite{Valerie01,Scuiller96} which we label as CrCu$_{6}$,
CrNi$_{6}$ and CrMn$_{6}$ respectively. In these molecules a Cr(III)
ion is surrounded by six cyanide ions, each bonded to a Cu(II), Ni(II)
or Mn(II) ion. The coordination sphere of Cr and Cu/Ni/Mn can be
described as a slightly distorted octahedral. The objective in this
work is to find the spin-spin correlation time $\tau (T)$ in the three
systems, and to compare them. Our main findings are: (I) $\tau $ is
nearly $T$ independent at low temperatures, and (II) at very low
temperature $\tau $ does not depend on $S$ or $J$ in this isotropic
case.

\begin{figure}[h]
\centerline{\epsfysize=6.0cm \epsfbox{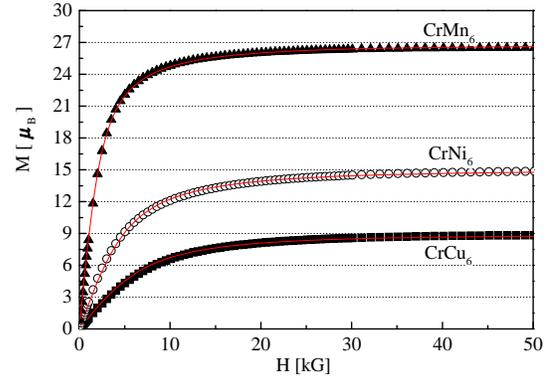}}
\caption{The magnetization of CrCu$_6$, CrNi$_{6}$ and CrMn$_{6}$ as a
function of the external applied field at $T=2$ K, respectively. The solid
lines are the $S=9/2$, $S=15/2$ and $S=27/2$ Brillouin functions (see text).}
\label{mag}
\end{figure}
In Fig.~\ref{mag} we show the magnetization $M$ per molecule in units
of $\mu _{B}$ as a function of applied field $H$ for the three
molecules. The data are taken at $T=2$ K. In all cases $M$ increases
as the applied field is increased. The magnetization reaches a
saturation value of $\frac{9}{2}$ $g \mu_{B}$, $\frac{15}{2}$ $g \mu_{B}$
and $\frac{27}{2}$ $g\mu _{B}$ in CrCu$ _{6}$, CrNi$_{6}$ and
CrMn$_{6}$ respectively. These saturation values are consistent with
$6$ Cu ($S=1/2$), Ni ($S=1$) or Mn ($S=5/2$) ions ferromagnetically
(or anti-ferromagnetically in the case of CrMn$_{6}$) coupled only to
a Cr ion of spin $3/2$. The solid lines are Brillouin functions of the
respective $S$. In Refs \cite {Valerie01,Mallah95,Mallah96,Scuiller96}
the susceptibility was fitted to the one expected from the Heisenberg
Hamiltonian and the values $J_{{\rm CrCu}_{6}}=77$ K, $J_{{\rm
CrNi}_{6}}=24$ K and $J_{{\rm CrMn}_{6}}=-11$ K where
found. Therefore, the highest spin value of each molecule is well
seperated from other spin states (a few tens of degrees K). High field
ESR measurements (on CrNi$_{6}$) \cite{Barra} and susceptibility
measurements (on CrCu$_{6}$, CrNi$_{6}$ and CrMn$_{6}$) found no
evidence for anisotropy, namely, $D\simeq 0$. This is consistent with
the octahedral character of the molecules.

In our $\mu $SR experiments we measure the polarization $P(t,H)$ of a
positive muon spin implanted in the sample, as a function of time $t$
and magnetic field $H$, where $P(0,H)=1$. The field is applied in the
direction of the initial muon polarization. The positive muon decays
to a positron which is emitted in the direction of the muon spin, and
the polarization as a function of time is reconstructed by the
detection of the emitted positrons.

The measurements in all molecules are done at temperatures ranging
from 25~mK up to 300~K, and in fields ranging between zero and
$20$~kG. These experiments were performed at both ISIS and PSI,
exploiting the long time window in the first facility for slow
relaxation (high $T$), and the high time resolution in the second
facility for fast relaxation (low $T$).

\begin{figure}[h]
\centerline{\epsfysize=6.0cm \epsfbox{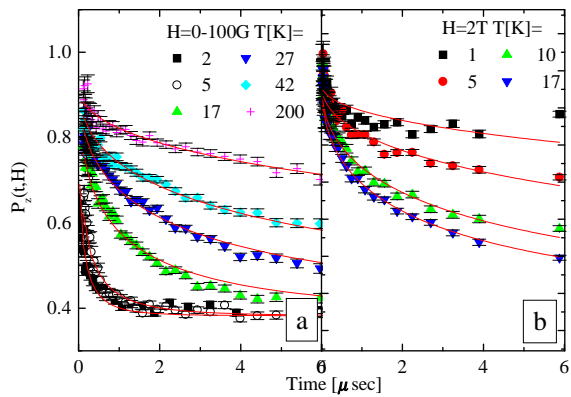}}
\caption{The spin polarization as a function of time. (a) At zero field and
different temperatures. (b) At field $H=20$~kG and different temperatures.
The solid lines are fits of the data to square root exponential functions.}
\label{asymmetry}
\end{figure}
In Fig.~\ref{asymmetry}(a) and (b) we present the muon spin
polarization as a function of time and for different temperatures in
CrNi$_{6}$ in zero field, and in $H=20$~kG, respectively. In zero
field, the relaxation rate increases, as the temperature is decreased,
and saturates at $\sim 5$~K. The increase at high temperatures is
caused by thermally activated tarnsitions between excited spin
states. However, at low temperatures, only the ground spin state is
populated, and only transitions within the degenerate ground state are
possible. In contrast, at $H=20$~kG, and temperatures lower than $\sim
17$~K, the relaxation rate decreases as the temperature is decreased,
and does not saturate.

In Ref.~\cite{Zaher2000} we demonstrated that the magnetic field experienced
by the muon in all molecules is dynamically fluctuating even at $T=50$~mK.
We therefore analyze our data using spin lattice relaxation theory. In this
theory, the polarization of a local probe (muon or nucleus), in the fast
fluctuation limit, is given by 
\begin{equation}
P(H,t)=\left( P_{0}-P_{\infty }\right) \exp \left[ -t/T_{1}\right]
+P_{\infty }  \label{PHt}
\end{equation}
where $P_{0}$ is the initial polarization, $P_{\infty }$ is the equilibrium
polarization \cite{UemuraPRB85}, 
\begin{mathletters}
\label{T1vsH}
\begin{eqnarray}
T_{1}(H) &=&A+BH^{2},  \label{T1vsHa} \\
A &=&\frac{1}{\Delta ^{2}\tau },  \label{T1vsHb} \\
B &=&\frac{\gamma ^{2}\tau }{\Delta ^{2}},  \label{T1vsHc}
\end{eqnarray}
and $\gamma $ is the probe gyromagnetic ratio. The correlation time $\tau $
and mean square of the transverse field distribution at the probe site in
frequency units $\Delta ^{2}$ are defined by 
\end{mathletters}
\begin{equation}  \label{correlation}
\gamma ^{2}\left\langle {\bf B}_{\bot }(t){\bf B\bot }(0)\right\rangle
=\Delta ^{2}\exp \left( -t/\tau \right) .  \label{qtau}
\end{equation}
The fast fluctuation limit is obeyed when $\tau \Delta \ll 1$.

In $\mu $SR $P_{\infty }=0$, and the muon could occupy many different sites
in the sample, because the molecules are fairly large ($\sim 15$ \AA \
diameter) and are embedded in an organic surrounding. As a result one must
average over $\Delta $. Using the distribution \cite{UemuraPRB85} 
\begin{equation}
\rho (\Delta )=\sqrt{\frac{1}{2\pi }}\frac{\Delta ^{\ast }}{\Delta ^{2}}\exp
\left( -\frac{1}{8}\left[ \frac{\Delta ^{\ast }}{\Delta }\right] ^{2}\right)
,  \label{DisDelta}
\end{equation}
and allowing for a constant background ($B_{g}$) due to muons stopping
outside the sample one obtains 
\begin{equation}
P(t)=P_{0}\exp \left( -\sqrt{t/T_{1}^{\mu }}\right) +B_{g},  \label{fits}
\end{equation}
where $1/T_{1}^{\mu }$ is the muon spin relaxation rate. This form is in
agreement with the experimental results (see below). In addition, Eq.~\ref
{T1vsHa} still holds, while in Eq.~\ref{T1vsHb} and \ref{T1vsHc}, $\Delta $
is replaced by $\Delta ^{\ast }$.

\begin{figure}[h]
\centerline{\epsfysize=6.0cm \epsfbox{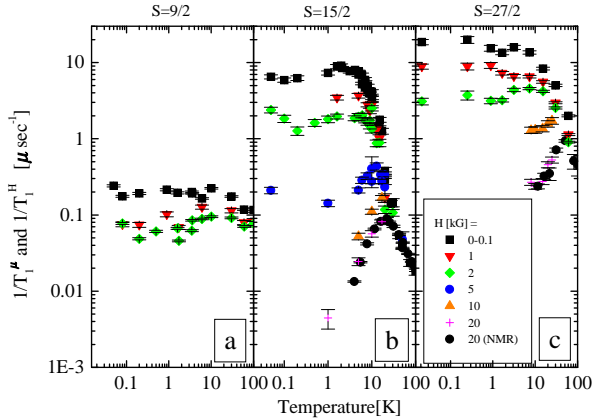}}
\caption{$1/T_{1}$ as a function of temperatures for different external
fields, measured by $\protect\mu $SR and by NMR (after scaling) in (a) CrCu$
_6$ (b) CrNi$_{6}$ and (b) CrMn$_{6}$.}
\label{lamvsT}
\end{figure}

The solid lines in Fig.~\ref{asymmetry} are fits of the data to Eq.~\ref
{fits} where $P_{0}$ is a global parameter. The parameter $B_{g}$ is free
within 10\% of its mean value since the high fields affect the positron
trajectory in a manner that is reflected in $B_{g}$. The fit is satisfactory
in all cases apart from the highest $H$ and lowest $T$ in CrNi$_{6}$. The
relaxation rate $1/T_{1}^{\mu }$ in the different compounds, obtained from
the fits is presented in Fig.~\ref{lamvsT} as a function of temperature for
different values of $H$. As pointed out above, at low fields $1/T_{1}^{\mu }$
increases with decreasing temperatures, and saturates at low temperatures.
In addition the value of $1/T_{1}^{\mu }$ increases as the spin of the
compound is higher, as expected from Eqs.~\ref{T1vsH} and \ref{correlation},
and the saturation temperature increases as the coupling constant $J$
increases. This is in strong contrast to Mn$_{12}$ \cite
{LascialfariPRL98LascialfariPRB98} and Fe$_{8}$ \cite{LascialfariPB00} where
in zero field $1/T_{1}^{\mu }$ increases continuously upon cooling until the
correlation time $\tau $ becomes so long that the molecule appears static in
the muon dynamical window, and no $T_1$ saturation is observed at low
temperatures.

\begin{figure}[h]
\centerline{\epsfysize=6.0cm \epsfbox{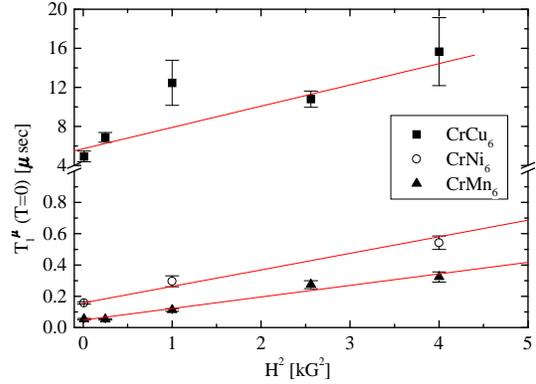}}
\caption{The saturation relaxation time as a function of $H^{2}$ for CrCu$_6$
, CrNi$_{6}$ and CrMn$_{6}$. The solid lines are linear fits of the data.}
\label{T1sat}
\end{figure}

In Fig.~\ref{T1sat} we plot the average relaxation time $T_{1}^{\mu
}(H)$ at $T\rightarrow 0$ as a function of $H^{2}$ for all compounds,
for fields up to $2$~kG (note the axis break). We find that
$T_{1}^{\mu }$ obeys Eq.~\ref {T1vsH}. This implies that the muon spin
relaxation is indeed due to dynamical field fluctuations, and that at
low $T$ the field autocorrelation can be described by a single
correlation time as long as the applied field is not too strong. At
high fields ($>$ 2 kG) we find deviations (not shown) from the linear
relation between $T_{1}$ and $H^{2}$. The deviation might be due to
the impact of the field on the spin dynamics, {\it i.e.} the
correlation function given by Eq.~\ref{qtau}. From the linear fits in
Fig.~\ref{T1sat}, and taking $(\Delta ^{\ast })^{2}=\gamma _{\mu
}(AB)^{-1/2}$ from Eq.~\ref{T1vsHb} and \ref{T1vsHc}, where $\gamma
_{\mu }=85.162$ ~MHz/kG, we find that for CrCu$_6$ $\Delta_{0}^{\ast
}=4.9\pm 0.9$~MHz ($57 \pm 10$ G), for CrNi$_{6}$ $\Delta _{0}^{\ast
}=26\pm 2$~MHz ($305\pm 25$ G), and for CrMn$_{6}$ $\Delta _{0}^{\ast
}=38\pm 2$~MHz ($446 \pm 24$ G); the subscript $0$ stands for
$T\rightarrow 0$. Using $\tau =(B/A\gamma _{\mu }^{2})^{1/2}$ from the
same equation we find $\tau _{0}=7\pm 1$~nsec for CrCu$_6$, $\tau
_{0}=10\pm 1$~nsec for CrNi$_{6}$ and $\tau _{0}=11\pm 1$~nsec for
CrMn$_{6} $. These values of $\Delta _{0}^{\ast }$ and $\tau _{0}$ are
self consistent with the fast fluctuation limit. Most striking is the
fact that all $\tau _{0}$ values are nearly equal.

Although our data support a picture where the muon spin relaxes due to
dynamically fluctuating magnetic fields, they leave open the
interpretation of these fluctuations. Are they due to the fluctuations
of the molecular spins or a result of muon diffusion, muonium
formation, etc.? In order to address this question we performed proton
NMR $T_{1}$ measurements. We find that in all applied fields smaller
than $20$~kG the proton $T_{1}$ is shorter than the experimental
window around the peak in $1/T_{1}^{\mu }$.  Only in a field of $\sim
20$~kG were we able to perform the experiment at all temperatures. We
measure $T_{1}$ using a saturation$-t-\pi /2-\pi $ pulse sequence. The
proton polarization recovery follows Eq.~\ref{PHt} with $P_{0}=0$ from
which we obtained $T_{1}^{H}$.

Since the proton gyromagnetic ratio $\gamma _{H}=42.57$ MHz/kG is very
different from that of the muon, we scale the NMR results and plot them
together with the $\mu $SR results in Fig.~\ref{lamvsT}, for CrNi$_{6}$ (b)
and CrMn$_{6}$ (c). The scaling factor $C$ used in Fig.~\ref{lamvsT} is $0.6$
for CrNi$_{6}$ and $8.8$ for CrMn$_{6}$. It was chosen so that the scaled
NMR relaxation rates will agree with the $\mu $SR rates at high $T$. After
this scaling, we find a good agreement between the $\mu $SR and NMR data at
all temperatures. In fields lower than $20$~kG, where the NMR $T_{1}$ was
measurable only at high $T$, we obtained the same agreement between the two
techniques.

The scaling factor provides information on the ratio of the field
experienced by a muon and by a proton. At high temperatures where
$T_{1}$ shows no field dependence one can assume that $A\gg BH^{2}$ in
Eq.~\ref {T1vsH}, and therefore $1/T_{1}=\Delta ^{2}\tau $. Using the
definition of $\Delta ^{2}$ given in Eq.~\ref{qtau} we can write
$C\equiv \left( \gamma _{H}/\gamma _{\mu }\right) ^{2}\left(
T_{1}^{H}/T_{1}^{\mu }\right) =\left\langle {\bf B}_{\perp
}^{2}\right\rangle ^{\mu }/\left\langle {\bf B}_{\perp
}^{2}\right\rangle ^{H}$ where $\left\langle {\bf B}_{\perp
}^{2}\right\rangle ^{H}$ is the RMS of the transverse field at the
proton site, and $\left\langle {\bf B}_{\perp }^{2}\right\rangle ^{\mu
}$ is the RMS of the field at the muon site in its general sense given
by Eq.~\ref {DisDelta}. The proximity of $C$ to 1 especially in the
CrNi$_{6}$ is very encouraging and suggests that the muon sites are
close to a proton in this system. Thus, we prove that in both
techniques we are measuring the probe's spin lattice relaxation time
due to the molecular spin fluctuations.

Finally, we would like to obtain the correlation time at all temperatures.
This could be calculated from $T_{1}^{\mu }$ combined with magnetization
measurements, at zero (or very low) fields. In zero order approximation we
assume that $\left\langle {\bf B}_{\perp }^{2}\right\rangle $ is
proportional to $\left\langle {\bf S}^{2}\right\rangle $ which is different
from $S(S+1)$ since at temperatures $k_{B}T\sim J$, states other than the
ground state $S$ can be populated. Therefore in zero field 
\begin{equation}
\frac{1}{T_{1}}(T)=\frac{\Delta_{0}^{\ast 2}\left\langle S^{2}\right\rangle
(T)\tau(T)}{S(S+1)}.
\end{equation}
Taking 
\begin{equation}
\left\langle S^{2}\right\rangle (T)=\frac{3k_{B}T\chi (T)}{N(g\mu _{B})^{2}},
\label{DCSusc}
\end{equation}
where $N$ is the number of molecules, $\mu _{B}$ is Bohr magneton, $k_{B}$
is the Boltzman factor, and $g=2$, we find 
\begin{equation}
\tau (T)=\frac{(g\mu _{B})^{2}}{3k_{B}}\frac{S(S+1)}{T_{1}(T)T\chi (T)\Delta
_{0}^{\ast 2}}.  \label{ratio}
\end{equation}

\begin{figure}[h]
\centerline{\epsfysize=6.0cm \epsfbox{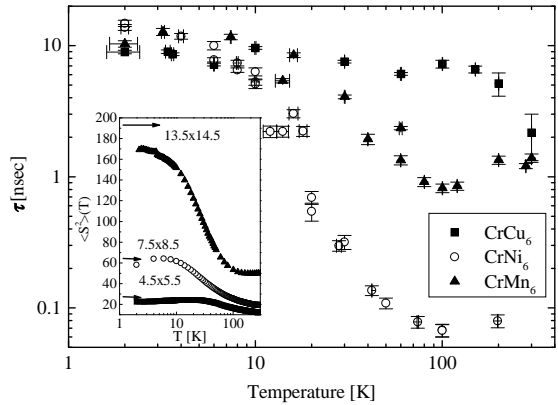}}
\caption{The correlation time $\protect\tau $ as given by Eq.~\ref{ratio} as
a function of temperature for CrCu$_6$, CrNi$_{6}$ and CrMn$_{6}$. The inset
shows $\left\langle S^{2}\right\rangle (T)$ as obtained from Eq.~\ref{DCSusc}
}
\label{tauT}
\end{figure}

In the insets of Fig.~\ref{tauT} we present $\left\langle
S^{2}\right\rangle (T)$, obtained from $H\rightarrow 0$
DC-susceptibility measurements and Eq.~\ref{DCSusc}, as a function of
temperature for CrCu$_6$, CrNi$_{6}$ and CrMn$_{6}$. In
Fig.~\ref{tauT} we present $\tau (T)$ as calculated using Eq.~\ref
{ratio} for the different compounds. The $T$ dependence of the
correlation time $\tau (T)$, unlike the muon spin lattice relaxation
rate, reflects the dynamics of the molecular spin without the $T$
dependence of the field at the muon site. At $T\sim 100$~K there is
more than an order of magnitude difference in $\tau $ between the
different molecules. As the temperature is lowered the correlation
time in all compounds increases as the temperature is decreased, but
reaches a {\em common saturation value} of $\sim 10$~nsec (within
experimental error) at $T\sim 10$~K. At this temperature only the
ground state $S$ is populated. In other words, when the HSM are formed
they all have the same correlation time at low $T$.

At very low temperatures one can interpret the correlation time $\tau
$ in terms of a broadening of the spin levels, due to interactions
between a molecular spins and other molecular spins, or the
enviroment. This interaction is considered as a purturbation, the
strength of which should be of the order of $h\tau ^{-1}$. Therefore,
our results have two major indications.  First, the broadening in the
$T\rightarrow 0$ limit is not due to phonons since these die out
exponentially with temperature at $0.05-1$~K. Second, the broadening
cannot be explained by interactions which are quadratic in $S$ or have
higher $S$ dependence. This rules out the dipolar interaction between
neighboring molecules since in the three compounds the nearest
neighbor distance is $\sim 15$ \AA. Similarly, crystal field terms
which are allowed by the octahedral symmetry ($S^4$ or higher
\cite{Abragam86}) are unlikely.

The only mechanisem suggested to date for level broadening of HSM, which is
weakly (linear) $S$ dependent is the hyperfine interaction between nuclear
and electronic spins. This mechanism can account for the finite spin
lattice relaxation rate at very low temperatures \cite{Waugh88PRB}. The
hyperfine interactions in unisotropic high spin molecules were studied
recently \cite{Caneschi98JMMM,Wernsdorfer99S}, and their effect on QTM is
becoming clearer \cite{Giraud01PRL,Giraud01condmat}. We believe that this
interaction also governs the fluctuations of the isotropic molecules at very
low temperatures ($T<3$ K). However, at high temperature ($T>10$ K) the
fluctuations are governed by spin-phonon interactions \cite{Salman01tobe}.

We would like to thank the ISIS and PSI muon facilities for their kind
hospitality and B.~Barbara for helpful discussion. These experiments were
supported by the European Union through its TMR Program for Large Scale
Facilities, the French Israeli cooperation program AFIRST, and the Israeli
Ministry of Science.


\begin{references}
\bibitem{ThomasNature96FriedmanPRL96}  L.~Thomas, F.~Lionti, R.~Ballou,
D.~Gatteschi, R.~Sessoli, and B.~Barbara, {\it Nature} {\bf 383}, 145
(1996). J.~R. Friedman, M.~P. Sarachik, J.~Tejada, and R.~Ziolo, {\it Phys.
Rev. Lett.} {\bf 76}, 3830 (1996).

\bibitem{SangregorioPRL97}  C.~Sangregorio, T.~Ohm, C.~Paulsen, R.~Sessoli,
and D.~Gatteschi, {\it Phys. Rev. Lett.} {\bf 78}, 4645 (1997).

\bibitem{AubinJACS98}  S.~M.~J.~Aunin, N.~R.~Dilley, M.~W.~Wemple,
M.~B.~Maple, G.~Christou, and D.~N.~Hendrickson, {\it J. Am. Chem. Soc.} 
{\bf 120}, 839 (1998).

\bibitem{GaraninPRB97PolitiPRL95} D.~A.~Garanin and E.~M.~Chudnovsky,
{\it Phys. Rev. B } {\bf 56}, 11102 (1997). P.~Politi, A.~Rettori,
F.~Hartmann-Boutron, and J.~Villain, {\it Phys. Rev. Lett.} {\bf 75},
537 (1995).

\bibitem{ProkofevPRL98}  N.~V. Prokof'ev and P.~C.~E. Stamp, {\it Phys. Rev.
Lett.} {\bf 80}, 5794 (1998).

\bibitem{GargPRL98}  A. Garg, {\it Phys. Rev. Lett.} {\bf 81}, 1513 (1998).

\bibitem{FriedmanJAP97}  J.~R.~Friedman, M.~P.~Sarachik, J.~M.~Hernandez,
X.~X.~Zhang, J.~Tejada, E.~Molins and R. Ziolo, {\it J. Appl. Phys.} {\bf 81}
, 3978 (1997)

\bibitem{VillainEL94}  J. Villain, F. Hartman-Boutron, R. Sessoli, and A.
Rettori, {\it Europhys. Lett.} {\bf 27}, 159 (1994).

\bibitem{PablosPRB97}  D. Garc\'ia-Pablos, N. Garc\'ia and H. De Raedt, {\it 
Phys. Rev. B} {\bf 55}, 931 (1997).


\bibitem{Valerie01}  V. Marvaud, private communication.

\bibitem{Mallah95}  T. Mallah, C. Auberger, M. Verdaguer and P. Veillet, 
{\it J. Chem. Soc., Chem. Commun.}, 61 (1995).

\bibitem{Mallah96}  T. Mallah, S. Ferlay, A. Scuiller and M. Verdaguer, {\it 
Molecular Magnetism: a Supramolecular Function}, NATO ASI series, C474,
Reidel, Dordrecht, 597 (1996).

\bibitem{Scuiller96}  A. Scuiller, T. Mallah, M. Verdaguer, A. Nivorozkhin,
J. L. Tholence and P. Veillet, {\it New J. Chem.} {\bf 20}, 1 (1996).

\bibitem{Barra}  A. Barra and M. Verdaguer, Private communication.


\bibitem{Zaher2000}  Z. Salman, A. Keren, P. Mendels, A. Scuiller and M.
Verdaguer, {\it Physica B} {\bf 289-290}, 106 (2000).

\bibitem{UemuraPRB85}  Y.~J.~Uemura, T.~Yamazaki, D.~R.~Harshman, M.~Senba,
E.~J.~Ansaldo, {\it Phys. Rev. B} {\bf 31}, 546 (1985). A. Keren, {\it Phys.
Rev. B} {\bf 50}, 10039 (1994).

\bibitem{LascialfariPRL98LascialfariPRB98}  A. Lascialfari, Z. Jang, F.
Borsa, P. Carretta, and D. Gatteschi, {\it Phys. Rev. Lett} {\bf 81}, 3773
(1998). A. Lascialfari, D. Gatteschi, F. Borsa, A. Shastri, Z. Jang, and P.
Carretta, {\it Phys. Rev. B} {\bf 57}, 514 (1998)

\bibitem{LascialfariPB00}  A. Lascialfari, P. Carretta, D. Gatteschi, C.
Sangregorio, J. Lord, and C. Scott, {\it Physica B} {\bf 289-290}, 110
(2000).

\bibitem{Abragam86} A. Abragam, B. Bleaney, {\it Electron Paramagnetic Resonance o f Transition Ions}; Dover Publications: Mineola, 1986.

\bibitem{Waugh88PRB}  J.~S.~Waugh and C.~P.~Slichter, {\it Phys. Rev. B} 
{\bf 37}, 4337 (1988).

\bibitem{Salman01tobe}  Z. Salman, to be published.

\bibitem{Caneschi98JMMM}  A.~Caneschi, D.~Gatteschi, C.~Sangregorio,
R.~Sessoli, L.~Sorace, A.~Cornia, M.~A.~Novak, C.~Paulsen and
W.~Wernsdorfer, {\it JMMM} {\bf 200}, 182 (1999).

\bibitem{Wernsdorfer99S}  W. Wernsdorfer and R. Sessoli, {\it Science} {\bf 
284}, 133 (1999).

\bibitem{Giraud01PRL}  R. Giraud, W. Wernsdorfer, A. M. Tkachuk, D. Mailly,
and B. Barbara, {\it Phys. Rev. B} {\bf 87}, 57203 (2001).

\bibitem{Giraud01condmat}  R. Giraud, W. Wernsdorfer, A. M. Tkachuk, D.
Mailly, and B. Barbara, {\it cond-mat/0108133}.
\end{references}
\end{document}